\documentclass[prl,twocolumn,aps]{revtex4-1}
\usepackage{epsfig}
\usepackage[dvips]{color}
\newcommand{\B}[1]{{\bm{#1}}}

\usepackage[latin1]{inputenc}
\newcommand{\beq}{\begin{equation}}
\newcommand{\eeq}{\end{equation}}
\newcommand{\bea}{\begin{eqnarray}}
\newcommand{\eea}{\end{eqnarray}}
\newcommand{\pa}{\partial}
\newcommand{\tr}{\hbox{tr}}

\usepackage{graphicx} 
\usepackage{subfigure}
\usepackage{dcolumn} 
\usepackage{bm} 
\usepackage{amssymb}

\begin{document}

\title{Dynamic stability of crack fronts: Out-of-plane corrugations}

\author{Mokhtar Adda-Bedia$^1$, Rodrigo E. Arias$^2$, Eran Bouchbinder$^3$ and Eytan Katzav$^4$}
\affiliation{$^1$Laboratoire de Physique Statistique, Ecole Normale Sup\'erieure,
UPMC Paris 6, Universit\'e Paris Diderot, CNRS, 24 rue Lhomond, 75005 Paris, France\\
$^2$Departamento de F\'isica, FCFM, University of Chile, Santiago, Chile\\
$^3$Chemical Physics Department, Weizmann Institute of Science, Rehovot 76100, Israel\\
$^4$Department of Mathematics, King's College London, Strand, London WC2R 2LS, UK}
\date{\today}
\begin{abstract}
The dynamics and stability of brittle cracks are not yet fully understood. Here we use the Willis-Movchan 3D linear perturbation formalism [J. Mech. Phys. Solids {\bf 45}, 591 (1997)] to study the out-of-plane stability of planar crack fronts in the framework of linear elastic fracture mechanics. We discuss a minimal scenario in which linearly unstable crack front
corrugations might emerge above a critical front propagation speed. We calculate this speed as a function of Poisson's ratio and show that corrugations propagate along the crack front at nearly the Rayleigh wave-speed. Finally, we hypothesize about a possible relation between such corrugations and the long-standing problem of crack branching.
\end{abstract}


\maketitle

Crack propagation is a major vehicle for material failure \cite{Freund, 10BMF}. Yet, a complete theoretical understanding of the dynamics and stability of rapidly propagating tensile cracks in brittle materials is lacking \cite{Ravi98, 99FM}. In the last few decades it has been experimentally established that rapid cracks undergo various types of dynamic instabilities. Planar crack fronts can experience a 3D micro-branching instability \cite{Ravi84a, Branching95, Branching96, Branching98}, a 2D oscillatory instability \cite{osc2007}, distributed micro-cracking in their vicinity \cite{Ravi84b, Ravi97, Scheibert2010, Guerra2012} and support propagating 3D waves, with both in-plane and out-of-plane components \cite{Sharon2001, Sharon2002, Fineberg2003, Ravi2003}.

Recently, progress in understanding these phenomena has been made. For example, linear in-plane crack front waves have been theoretically predicted \cite{Ramanathan1997, Ramanathan1997a, Morrissey1998, Morrissey2000}, dynamic crack branching has been observed in 2D numerical simulations \cite{numeric1, numeric2, numeric3, numeric4}, energetic bounds on 2D crack branching have been derived \cite{adda03, adda04b, adda04, adda05, katzav07} and the 2D oscillatory instability has been explained based on intrinsic nonlinearities \cite{bouchbinder2009, goldman2012}, whose role in dynamic fracture has been highlighted in \cite{Buehler2003, Buehler2006, 08LBF, 08BLF, 09BLF, 10BLIF, 10Bou}. While progress in understanding 3D crack front instabilities in the context of quasi-static mixed-mode fracture \cite{Pons2010, Leblond2011} and dynamic tensile fracture \cite{Willis2012} has been made very recently, the 3D out-of-plane nature of crack front waves \cite{Sharon2001} and the micro-branching instability \cite{99FM} remain theoretically elusive.

In this Letter we study the out-of-plane stability of planar crack fronts in the framework of linear elastic fracture mechanics using the Willis-Movchan 3D linear perturbation formalism \cite{Willis1995, Movchan1995, Willis1997, Willis1999, Willis2003}. The emergence of out-of-plane crack corrugations within a minimal scenario is discussed. We calculate the critical propagation speed for the onset of such corrugations, which propagate along the crack front at nearly the Rayleigh wave-speed, as a function of Poisson's ratio. We then speculate about a possible relation between these corrugations and the 3D micro-branching instability and its fractographic consequences. Our work is directly inspired by \cite{Willis2012}.

To set the stage for our stability analysis, consider a planar crack propagating steadily in an isotropic linear elastic solid and introduce small perturbations which give rise to the following perturbed crack surface configuration
\bea
\label{perturbed}
&&S_\epsilon(t)=\{{\bm x}\!:\!-\infty\!<\!x_1\!<\!vt+\epsilon\,\phi(x_2,t),\nonumber\\
&&-\infty\!<\!x_2\!<\!\infty, \,x_3\!=\!\epsilon\,\psi(x_1,x_2,t)\} \ ,
\eea
where $\{x_1,x_2,x_3\}$ is a fixed Cartesian coordinate system and $v$ is a constant propagation speed in the $x_1$-direction, smaller than the Rayleigh wave-speed $c_R$.
The smooth and bounded functions $\phi$ and $\psi$ describe in-plane and out-of-plane perturbations, respectively, and $0 \!<\!\epsilon \!\ll\! 1$ is a small dimensionless amplitude. Note that in Eq. (\ref{perturbed}) we use $\infty$ to schematically represent finite macroscopic length scales that are much larger than any length scale characterizing the perturbations. Such scales will be explicitly invoked below and will play a role in our analysis.

The isotropic linear elastic solid is characterized by a stress field $\B\sigma$ and a displacement field $\B u$ that satisfy the momentum balance equation $\nabla\!\cdot\B\sigma\!=\!\rho \pa_{tt}\B u$, where $\rho$ is the mass density, and are related through Hooke's law $\B\sigma\!=\!\lambda \,\tr\B\varepsilon\,\B I \!+\! 2\mu \,\B\varepsilon$, where $\B\varepsilon\!=\!\case{1}{2}[\nabla \B u+ (\nabla \B u)^T]$. Here $\B I$ is the identity tensor, and $\lambda$ and $\mu$ are the Lam\'e constants that are related to the dilatational and shear wave-speeds through $c_d\!=\!\sqrt{(\lambda+2\mu)/\rho}$ and $c_s\!=\!\sqrt{\mu/\rho}$, respectively. The crack surfaces are traction-free $\sigma_{ij}n_j\!=\!0$, where $i\!=\!1\!-\!3$ and $\B n$ is an outward unit normal to $S_\epsilon(t)$, and some external conditions are specified on the outer boundaries of the solid.
This formulation constitutes Linear Elastic Fracture Mechanics (LEFM) \cite{Freund}.

A fundamental prediction of this framework is that the asymptotic crack front expansion of the stress field $\B\sigma$ is dominated by a universal square-root singularity in the distance from the front. For the unperturbed crack (Eq. (\ref{perturbed}) with $\epsilon\!=\!0$), under pure (symmetric) tensile loading conditions, the expansion of the tensile stress component ahead of the front takes the form \cite{Freund}
\beq
\label{expansion}
\sigma^{(0)}_{33} \simeq K^{(0)}_I \left(2\pi X\right)^{-1/2} + A^{(0)}_3 X^{1/2}\quad\hbox{for}\quad X \!\to\! 0^+ \ ,
\eeq
where $X \!=\! x_1-v t$. Here $K^{(0)}_I$ is the mode I (tensile) Stress Intensity Factor (SIF), a fundamental quantity that quantifies the intensity of the linear elastic singularity, $A^{(0)}_3$ is the coefficient of the sub-leading term to be used below and the superscript $(0)$ refers to the unperturbed crack. Note that $K^{(0)}_I$ and $A^{(0)}_3$ are assumed to be independent of $x_2$ and $t$. The crack front perturbations described in Eq. (\ref{perturbed}) break both the translational symmetry along $x_2$ and the $x_3\!\to\!-x_3$ symmetry of the global loading. Therefore, the stress $\B\sigma$ associated with the perturbed crack admits the same asymptotic expansion (ahead of the front, along its local tangent) as in Eq. (\ref{expansion}), but now $K_I(x_2,t)$ depends on $x_2$ and $t$, and there exist also mode II (in-plane shear) and mode III (tearing) singular contributions proportional to $K_{II}(x_2,t)$ and $K_{III}(x_2,t)$, respectively.

The major goal of a linear perturbation theory is to calculate the SIF's $K_{I,II,III}$ to linear order in $\epsilon$:
$K_j(x_2,t)\!=\!K^{(0)}_I\delta_{jI}\!+\!\epsilon K^{(1)}_j(x_2,t)$, with $j\!=\!I,II,III$. This problem was solved by Willis and Movchan in 1997, yielding \cite{Willis1997}
\begin{eqnarray}
\label{KI,II}
K_{I}^{(1)}(x_2,t) &=& Q_{33}\!\ast\!\phi K_I^{(0)}\!+\!\sqrt{\pi/2}\,\phi A_3^{(0)} \ ,\\
\label{KI,IIa}
K_{II}^{(1)}(x_2,t) &=& \left[ -\Theta_{13}Q_{11} \ast\psi^\ast  \!-\! \omega_{13}\pa_{x_1}\psi^\ast\right]  K_I^{(0)}\\
&-&\Theta_{13}\psi^ \ast A_3^{(0)}\!+\!T_{II}(x_2,t) \nonumber \ ,
\end{eqnarray}
where $\ast$ stands for a convolution in $(x_2,t)$ and $\psi^\ast$ is $\psi$ evaluated at the crack front. We do not present here $K_{III}$ because it plays no role in what follows.

Let us discuss the physical meaning of the various terms in Eqs. (\ref{KI,II})-(\ref{KI,IIa}). $Q_{ij}$ are weight-functions that quantify the non-local (both in space and time) elastodynamic interaction of the crack front with itself. They admit explicit Fourier transform representations $\overline{Q}_{ij}$, and are homogeneous functions of degree one (see \cite{Willis1997, Willis1999, Willis2003, Willis2012, supplementary} for details). $T_{II}(x_2,t)$ involves convolutions of additional weight-functions with the ``history'' of out-of-plane perturbations, i.e. $\psi$ for $X\!<\!0$, and effective tractions associated with zero order stresses \cite{Willis1997, supplementary}. The other terms are purely geometric in nature and correspond to local rotations and translations of the crack front. Finally, the kinematic functions $\Theta_{13}(v)$ and $\omega_{13}(v)$ take the form
\bea
\Theta_{13}(v)&=& -2\alpha(\beta-\alpha)(1+\beta^2)/R(v)\ , \\
\omega_{13}(v)&=&-[8\alpha \beta -(1+\beta^2)(2+\alpha^2+\alpha\beta)]/R(v) \ ,
\eea
where $\alpha^2\!=\!1-v^2/c_d^2$, $\beta^2\!=\!1-v^2/c_s^2$ and $R(v)\!=\!4\alpha\beta-(1+\beta^2)^2$ (whose root is the Rayleigh wave-speed $c_R$).

In order to prepare Eqs. (\ref{KI,II})-(\ref{KI,IIa}) for the stability analysis to follow, we first Fourier transform all relevant functions with respect to $(x_2,t)$ such that $\overline{f}(\xi_2,\omega)$ is the transform of $f(x_2,t)$. Noting that to leading order in $\epsilon$ one has $\pa_{x_1}\psi^\ast\!=v^{-1}\!\pa_t\psi^\ast$, the wave-number of the perturbation $\psi$ in the propagation direction $x_1$ is readily given by $\xi_1\!=\!\omega/v$. In a minimal scenario, a crack problem features one macroscopic length scale $L$ and one macroscopic stress scale $\sigma^\infty$. Dimensional analysis then implies that $K^{(0)}_I \!\sim\! \sigma^\infty \sqrt{L}$ and $A^{(0)}_3\!=\!K^{(0)}_I a_3(v)/L$, where $a_3(v)$ is a real dimensionless function. Furthermore, the Fourier transform of $T_{II}(x_2,t)$ can be expressed as
\bea
\hspace{-0.4cm}\overline{T}_{II}(\xi_2,\omega)\!=\! K_I^{(0)}\sqrt{|\xi_2|/L}\,\overline{t}_{II}(\omega/\left|\xi_2\right|,\left|\xi_2\right|L)\overline{\psi^*}(\xi_2,\omega),
\eea
where $\overline{t}_{II}$ is a complex function.

Therefore, the Fourier space version of Eqs. (\ref{KI,II})-(\ref{KI,IIa}) reads
\begin{eqnarray}
\label{KIIf}
&&\overline{K}_{II}^{(1)}\!(\xi_2,\omega) = \Big[\!-\!\left|\xi_2\right|\Theta_{13}\overline{q}_{11}(\omega/\left|\xi_2\right|)
 +i(\omega/v)\,\omega_{13}\\ &&-\Theta_{13} a_3/L+\sqrt{|\xi_2|/L}\,\overline{t}_{II}(\omega/\left|\xi_2\right|,\left|\xi_2\right|L)\Big]  \overline{\psi^*}(\xi_2,\omega) K_I^{(0)} \ , \nonumber\\
\label{KIf}
&&\overline{K}_{I}^{(1)}\!(\xi_2,\omega)\!=\!\left[\left|\xi_2\right|\overline{q}_{33}\!(\omega/\left|\xi_2\right|)\!+\!\sqrt{\frac{\pi}{2}}\frac{a_3}{L}\right]\!\overline{\phi}(\xi_2,\omega) K_I^{(0)},
\end{eqnarray}
where we used the homogeneity property of $\overline{\B Q}$ to define $\overline{Q}_{11,33}(\xi_2,\omega)\!=\!\left|\xi_2\right|\overline{q}_{11,33}(\omega/\left|\xi_2\right|)$. Equations (\ref{KIIf})-(\ref{KIf}), which serve as a basis for our stability analysis, show that in-plane and out-of-plane perturbations are decoupled to first order in $\epsilon$ \cite{Willis1997}, though experiments demonstrate such a coupling \cite{Sharon2001}, suggesting the intervention of higher order corrections.

In order to study the out-of-plane stability of the crack, we need an equation of motion for the front in terms of the SIF's. It is well-established that under quasi-static and 2D (i.e. when the front can be treated as a tip) conditions, the Principle of Local Symmetry (PLS) $K_{II}\!=\!0$ \cite{Goldstein}, is valid \cite{Pham2008, Corson2009}. This implies that no intrinsic nonlinear scales near the crack front play a role in crack instabilities and hence that crack patterns are always determined by extrinsic (typically geometric) length scales. However, it was recently shown that for strongly dynamic conditions (i.e. $v$ of the order of $c_R$) in 2D, a generalization of the PLS is required and intrinsic nonlinear scales play a decisive role in rapid crack instabilities \cite{bouchbinder2009, goldman2012}. The role of intrinsic length scales in the context of quasi-static 3D crack propagation under mixed-mode $K_I+K_{III}$ conditions has also been demonstrated recently \cite{Pons2010, Leblond2011}.

Here, however, we seek a minimal scenario for the emergence of out-of-plane instabilities and hence wish to deviate as little as possible from LEFM. Therefore, we adopt the PLS, $K_{II}\!=\!0$, even though we consider 3D cracks under dynamic conditions. Finally, as a crack propagation condition we use the generalized Griffith criterion, quantifying the balance between energy flowing into the crack front region and the dissipation associated with the fracture process \cite{Freund}, which together with the PLS reads \cite{adda99}
\beq
\label{G}
\alpha(1-\beta^2)K_I^2/[2\mu R(v)]=\Gamma\quad\hbox{and}\quad K_{II} = 0  \ .
\label{eqmo}
\eeq
Here $\Gamma$ is the fracture energy, assumed to be $v$-independent. Note that in general, the energy balance relation contains also terms proportional to $K_{II}^2$ and $K_{III}^2$ \cite{Freund}; the former is omitted due to the $K_{II}\!=\!0$ condition and the latter because it does not contribute to order $\epsilon$.

\begin{figure}[ht]
\centering
\epsfig{width=0.55\textwidth,file=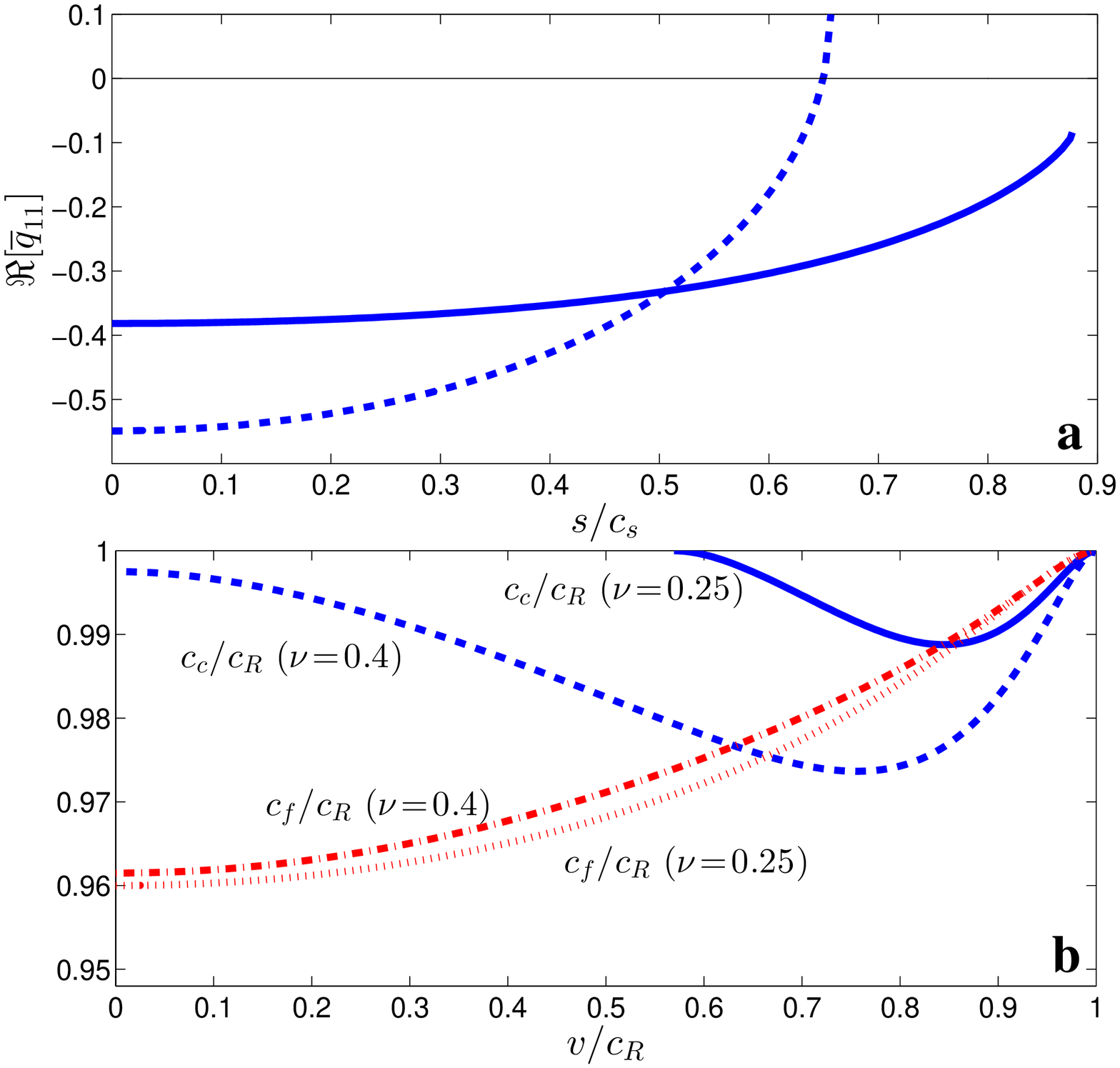}
\caption{(Color online) (a) $\Re\left[\overline{q}_{11}\right]$ vs. $s/c_s$ with $\nu\!=\!0.25$ for $v\!=\!0.3c_R$ (solid line, no zero-crossing) and $v\!=\!0.7c_R$ (dashed line, zero-crossing). (b) The normalized crack corrugation wave-speed $c_c/c_R$ vs. $v/c_R$ for $\nu\!=\!0.25$ (solid blue line) and $\nu\!=\!0.4$ (dashed blue line). For completeness, we show also the normalized in-plane crack front wave-speed $c_f/c_R$ for $\nu\!=\!0.25$ (dotted red line) and $\nu\!=\!0.4$ (dash-dotted red line).}\label{fig1}
\end{figure}

Before discussing the out-of-plane stability, we briefly consider the in-plane dynamics of the crack front \cite{Ramanathan1997a}. Expanding the energy balance relation in Eq. (\ref{G}) and using $\overline{K}_{I}^{(1)}\!(\xi_2,\omega)$ of Eq. (\ref{KIf}), we obtain
\begin{eqnarray}
\hspace{-0.7cm}2\,\overline{q}_{33}\!\left(\frac{\omega}{\left|\xi_2\right|}\right) \!-\!i\frac{\omega}{\left|\xi_2\right|}\pa_v\ln\!\left[\!\frac{\alpha(1-\beta^2)}{R(v)}\!\right]\!+\!\frac{\sqrt{2\pi}a_3}{\left|\xi_2\right|L} = 0 \label{eqI} \ .
\end{eqnarray}
In the limit $|\xi_2|L \!\gg\! 1$ this equation admits a real solution for every $v$, giving rise to a dispersion relation of the form $\omega\!/|\xi_2|\!=\!s_f(v)\!+\!O(1/|\xi_2|L)$, where $c_f^2\!=\!s_f^2+v^2$ is the propagation speed in a fixed coordinate system ($s_f$ is the propagation speed seen by an observer moving with the crack front). These are the well-known in-plane crack front waves \cite{Ramanathan1997, Ramanathan1997a, Morrissey1998, Morrissey2000}, whose speed $c_f$ is plotted in Fig. \ref{fig1}b. Note that while the linear perturbation theory determines a dispersion relation in terms of $\omega$ and $|\xi_2|$, it cannot select them separately. Moreover, although in-plane crack fronts might play a role in fracture dynamics, they cannot leave fractographic marks by themselves, which entails an out-of-plane component.

To study the out-of-plane stability, we apply the PLS to Eq. (\ref{KIIf}), i.e. set $\overline{K}_{II}^{(1)}\!=\!0$. The outcome is a complex equation for a complex variable $\omega/|\xi_2|$ and a real variable $|\xi_2|L$. We expect $\overline{K}_{II}^{(1)}\!=\!0$ to be equivalent to
\beq
\label{stab}
\!\Re\left[\,\omega/|\xi_2|\right]\!=\!h\!\left(v,|\xi_2|L\right),~~~\Im\left[\,\omega/|\xi_2|\right]\!=\!g\!\left(v,\left|\xi_2\right|L\right) \ ,
\eeq
where $h(\cdot)$ and $g(\cdot)$ are in principle calculable. The crack stability against out-of-plane perturbations is determined by the sign of $\Im[\omega/|\xi_2|]$. For $\Im[\omega/|\xi_2|]\!<\!0$ perturbations are attenuated and the crack is stable. For $\Im[\omega/|\xi_2|]\!>\!0$ perturbations are amplified and the crack is unstable, developing surface corrugations. Therefore, the instability threshold is determined by $\Im[\omega/|\xi_2|]\!=\!0$ \cite{Pham2008, Corson2009}. Before we perform this analysis, let us discuss two possible scenarios. In the first, we set $|\xi_2|L\!\to\!\infty$. Therefore, at the instability threshold, the second equation in (\ref{stab}) reads $g(v,\infty)\!=\!0$. The solution, which determines the critical speed $v_c$ for instability, is then substituted into the first equation in (\ref{stab}), yielding $\Re[\omega/|\xi_2|]\!=\!h(v_c,\infty)$, which is a dispersion relation for corrugation waves (similar to the one derived for in-plane front waves). In the second scenario, we keep $|\xi_2|L$ finite (albeit large) and obtain $g(v,|\xi_2|L)\!=\!0$ and $\Re[\omega/|\xi_2]\!=\!h(v,|\xi_2|L)$. Here both $\omega$ and $|\xi_2|$ are being selected, and the critical speed $v_c$ (if exists) is determined from a solvability condition. In this case, the wave-number in the propagation direction, $\xi_1\!=\!\omega/v$, is also selected and the complete crack surface topography is determined.

To test these scenarios we use the stability condition $\Im[\omega/|\xi_2|]\!=\!0$ to obtain (for $v\!<\!c_R$) \cite{supplementary}
\beq
2\Im\left[\overline{q}_{11}(s)\right] = s\,\pa_v\ln\!\left[\beta(1-\beta^2)/R(v)\right] \ ,
\eeq
where $s\!\equiv\!\omega/|\xi_2|$ is now real. Substituting this result in $\overline{K}_{II}^{(1)}\!=\!0$ (see Eq. (\ref{KIIf})), we obtain
\bea
\label{onset1}
\hspace{-0.4cm}&&\Re\left[\overline{q}_{11}(s)\right]=
 -\frac{a_3}{\left|\xi_2\right|L}+\frac{\Re\left[\overline{t}_{II}\left(s,\left|\xi_2\right|L\right)\right]}{\Theta_{13}\sqrt{\left|\xi_2\right|L}}\ ,\\
\label{onset2}
\hspace{-0.4cm}&&\hat{\omega}_{13}=-v\,\Im\left[\overline{t}_{II}\left(s,\left|\xi_2\right|L\right)\right]/(s\,\sqrt{\left|\xi_2\right|L}) \ ,
\eea
where $2\,\hat{\omega}_{13}\equiv2\omega_{13}-\Theta_{13}v\,\pa_v\ln\!\left[\beta(1-\beta^2)/R(v)\right]$.

To analyze these equations within the first scenario we set $|\xi_2|L\!\to\!\infty$ such that Eq. (\ref{onset2}) becomes $\hat{\omega}_{13}(v)\!=\!0$. This equation admits no solution as $\hat{\omega}_{13}(v)\!<\!0$ for every $v\!<\!c_R$. Indeed, it was recently demonstrated that within this scenario there exist no solutions with real $\omega/|\xi_2|$ and that out-of-plane perturbations experience attenuation for every $v$ \cite{Willis2012}. To test the second scenario, we propose an iterative scheme for solving Eqs. (\ref{onset1})-(\ref{onset2}). As macroscopic length scales are assumed to be much larger than any length scale characterizing the perturbations, to leading order we can set $|\xi_2|L\!\to\!\infty$ on the right-hand-side of Eq. (\ref{onset1}), obtaining $\Re\left[\overline{q}_{11}(s)\right]\!\simeq\!0$. If the latter admits a solution, then $s$ is determined to order $O(1/\sqrt{\left|\xi_2\right|L})$ and Eq. (\ref{onset2}) yields $\sqrt{\left|\xi_2\right|L}\!=\!-v\,\Im\left[\overline{t}_{II}\left(s,\infty\right)\right]/(s\,\hat{\omega}_{13})$, which determines $|\xi_2|$ and requires the calculation of $\overline{t}_{II}$. Once $\overline{t}_{II}$ is known, the steps can be iterated to produce higher order corrections in powers of $1/\sqrt{\left|\xi_2\right|L}$. Here, however, we do not calculate $\overline{t}_{II}$, which is a complex function that depends on the geometry and loading conditions of any particular crack problem, and hence truncate the iterative scheme at its first step, $\Re\left[\overline{q}_{11}(s)\right]\!\simeq\!0$, assuming that the next steps yield a solution for $|\xi_2|$.

\begin{figure}[ht]
\centering
\epsfig{width=0.38\textwidth,file=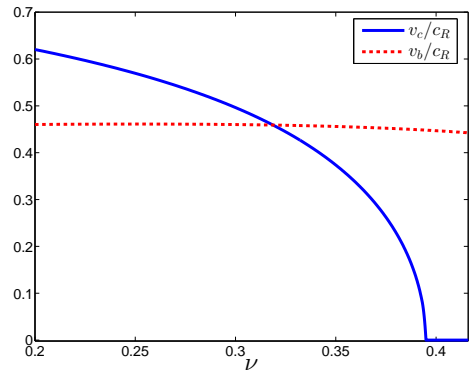}
\caption{(Color online) The normalized critical speed $v_c/c_R$ for the onset of out-of-plane instability (solid blue line) and the normalized critical speed $v_b/c_R$ above which 2D branching is energetically possible vs. $\nu$ (dashed red line) \cite{katzav07}.}\label{fig2}
\end{figure}

To explore this scenario we look for solutions of $\Re\left[\overline{q}_{11}(s)\right]\!=\!0$, which we denote by $s_c$. It turns out that a real solution $s_c$ exists only above a critical propagation speed $v_c$, which is a function of Poisson's ratio $\nu\!=\!\lambda/[2(\lambda\!+\!\mu)]$. Examples are shown in Fig. \ref{fig1}a. Once $s_c$ is known for a given $v$, the crack corrugation wave-speed $c_c$, in a fixed coordinate system, is given by $c_c^2\!=\!s_c^2+v^2$. In Fig. \ref{fig1}b $c_c/c_R$ is plotted vs. $v/c_R$ for two values of $\nu$ and $v\!\ge\!v_c$. $c_c$ is nearly the Rayleigh wave-speed $c_R$. For completeness, we plot also the in-plane crack front wave-speed $c_f$, which exists for every $v$. In Fig. \ref{fig2} we show the critical speed for the onset of out-of-plane corrugations $v_c$ as a function of Poisson's ration $\nu$. This constitutes our main result: the existence of an out-of-plane crack instability above a critical speed within a minimal scenario, accompanied by the propagation of corrugation waves along the crack front. $v_c$ is a monotonically decreasing function of $\nu$, which vanishes above $\nu_c\!\simeq\!0.4$.

Might the out-of-plane corrugations discussed here be related to the micro-branching instability? Providing a definite answer goes well beyond the scope of the present work. It is, however, conceivable that the two phenomena are related. A micro-branching event, like the corrugations discussed above, breaks the translational symmetry along $x_2$ (it is localized along this dimension \cite{Branching96}) and goes out-of-plane. One may then speculate that the out-of-plane front instability may trigger the micro-branching instability, or if the two instabilities are of different origins, that one may amplify/suppress the other \cite{Livne}.

To see what the implications of this speculation might be, we need an estimate for the critical speed for micro-branching $v_b$. In the absence of a better alternative, we use the energetic bound for 2D branching \cite{katzav07}, which is shown in Fig. \ref{fig2} together with $v_c$. The two curves intersect at $\nu\!\simeq\!0.32$. Consider materials with $\nu\!\lesssim\!0.32$ (e.g. $\nu\!\simeq\!0.2$ for silica glass) where $v_b\!<\!v_c$. For $v\!<\!v_b\!<\!v_c$ out-of-plane perturbations are stable and branching is not possible. For $v_b\!<\!v\!<\!v_c$ out-of-plane perturbations are still stable, though branching is energetically possible. Finally, for $v_b\!<\!v_c\!<\!v$ out-of-plane perturbations are unstable and branching is energetically possible. Consider then materials with $\nu\!\gtrsim\!0.32$ (e.g. $\nu\!\simeq\!0.35$ for Plexiglas or $\nu\!\simeq\!0.5$ for elastomer gels). Here $v_c\!<\!v_b$ and there exists a regime with $v_c\!<\!v\!<\!v_b$, where out-of-plane perturbations are unstable, while branching is energetically prohibited.

As corrugations and branching events leave traces on fracture surfaces, the different $v$ regimes described above might be related to the common mirror-mist-hackle transition, a fractographic characterization that corresponds to progressive increase in surface roughness with increasing crack speed \cite{Hull}. For example, in a regime where the crack front is stable against both corrugations and branching, we expect mirror-like surfaces. For $\nu\!>\!\nu_c\!\simeq\!0.4$, where such a regime does not exist, we do not expect mirror-like surfaces even at very low propagation speeds. In a regime where only one of the instabilities can be triggered, typically at higher $v$'s, surface roughness will emerge, which might correspond to the mist zone. When both instabilities can take place, at yet higher $v$'s, we expect denser surface features and enhanced roughness, which might correspond to the hackle zone. We hope to further explore these issues in the future.

While our analysis, assuming LEFM and the PLS, implies that the scale of out-of-plane corrugations is determined by an extrinsic (geometric or loading-inherited) length scale $L$, in practice such corrugations typically feature much smaller scales \cite{99FM, Hull}. Nevertheless, we believe our analysis remains relevant also when intrinsic length scales, associated with near crack front nonlinearities, microstructures and dissipation, play a role in selecting the scales of corrugations. The idea is that the SIF's uniquely couple the large scales of a crack problem to the small scales near the front.
Therefore, whenever a region where the square-root singular SIF's fields dominate the mechanical response exists, which is the case for a broad range of materials \cite{Freund}, the calculation of the SIF's in Eqs. (\ref{KIIf})-(\ref{KIf}) is required. Moreover, we expect any generalization of the PLS -- incorporating the role of intrinsic scales -- to be represented in terms similar to $\overline{t}_{II}$ and thus the structure of Eqs. (\ref{onset1})-(\ref{onset2}) to remain valid, where the role of $L$ is played now by an intrinsic length scale.

To conclude, we demonstrate -- within a minimal linear elastic fracture mechanics scenario -- the existence of an out-of-plane crack front instability characterized by a Poisson's ratio dependent critical propagation speed. We speculate on a possible relation between this instability and the 3D micro-branching instability, and its fractographic implications. Extending our analysis to explicitly include intrinsic nonlinear effects \cite{bouchbinder2009, goldman2012} and finite geometric scales is an important direction for future investigation.


We are grateful to J.R. Willis, N.V. Movchan and A.B. Movchan for generously sharing with us their results prior to publication.
E.B. is grateful to J. Fineberg for his continuous support and to A. Livne for enlightening discussions. E.B. acknowledges support from the James S. McDonnell Fund, the Minerva Foundation with funding from the Federal German Ministry for Education and Research, the Harold Perlman Family Foundation and the William Z. and Eda Bess Novick Young Scientist Fund. R.E.A and M.A.B acknowledge support from CNRS-Conicyt project 170.


\newpage
\onecolumngrid

\renewcommand{\theequation}{A\arabic{equation}}
\renewcommand{\thefigure}{S\arabic{figure}}
\vspace{2cm}
\begin{center}
{\large{\bf Supplementary Information}}
\end{center}

The aim of this document is to provide additional technical details about the calculations of the complex functions $\overline{q}_{33}(s)$  in Eq. (11) and $\overline{q}_{11}(s)$ in Eqs. (13)-(14) in the manuscript.

\section{First order perturbations of the stress intensity factor}

Willis and Movchan introduced a convolution identity (Eq. (3.12) of \cite{Willis1997_S}) that allows to calculate the change of the stress intensity factors due to a change in the crack surface, induced by a perturbation of its propagation front. Starting from an unperturbed state defined by a large crack submitted to a mode~I loading,
the first order perturbation terms of the stress intensity factors, $\overline{K}_{I}^{(1)}(\xi_2,\omega)$ and $\overline{K}_{II}^{(1)}(\xi_2,\omega)$ are given by~(Eqs. (3)-(4) of the main text)
\begin{eqnarray}
\overline{K}_{I}^{(1)}(\xi_2,\omega) & = & \left[\overline{Q}_{33}(\xi_2,\omega) K_I^{(0)}+\sqrt{\frac{\pi}{2}}A_3^{(0)}\right] \overline{\phi}(\xi_2,\omega)  \label{KIF}\ ,\\
\overline{K}_{II}^{(1)}(\xi_2,\omega) & = & \left[ -\Theta_{13}\overline{Q}_{11}(\xi_2,\omega)K_I^{(0)}
 +i(\omega/v) \omega_{13} K_I^{(0)} -\Theta_{13} A_3^{(0)} \right]  \overline{\psi}^*(\xi_2,\omega) + \overline{T}_{II}(\xi_2,\omega) \label{KIIF}\ .
\end{eqnarray}
Explicit expressions for the quantities in Eqs.~(\ref{KIF})-(\ref{KIIF}) can be found in~\cite{Willis1999_S}. The ``history'' term $T_{II}$ includes the contributions of the zero order traction terms on the broken surface convoluted with the corrugations of the crack surface
$\psi(X<0,x_2,t)$. It is expressed as~\cite{Willis1999_S}
\beq
T_{II}(x_2,t)=  \left[U\right]_{11}\ast \langle P^{(1)}_1\rangle +\left[U\right]_{21}\ast \langle P^{(1)}_2\rangle
-\langle U \rangle_{31} \ast [P_3^{(1)} ] \ ,
\eeq
where $[\cdot]$ and $\langle \cdot\rangle$ denote the jump and the average of the corresponding quantities through the unperturbed crack plane. Here $\mathbf{U}$ are the weight functions defined in \cite{Willis1999_S} and $\mathbf{P}^{(1)}$ are traction terms obtained by imposing the boundary conditions on the edges of the perturbed crack. The latter are given by
\begin{equation}
P_i^{(1)} =-\pa_\alpha\left(\sigma_{i\alpha}^{(0)}\psi\right)+\rho v^2 u_{i,XX}^{(0)} \psi \ .
\end{equation}
Also, one can show that $\overline{K}_{I}^{(1)}$ does not include such a term, i.e. that $T_I(x_2,t)=0$.

The functions $\overline{Q}_{ij}(\xi_2,\omega)$ are homogeneous of degree one and can be written as
\beq
\overline{Q}_{11}(\xi_2,\omega)=\left|\xi_2\right|\overline{q}_{11}(\omega/\left|\xi_2\right|),\qquad \overline{Q}_{33}(\xi_2,\omega)=\left|\xi_2\right|\overline{q}_{33}(\omega/\left|\xi_2\right|)\ ,
\eeq
where $\left|\xi_2\right|$ is the modulus of the {\it complex} wavenumber $\xi_2$ and $\omega$ is a {\it complex} frequency. The functions $\overline{q}_{11}$ and $\overline{q}_{33}$ are given in~\cite{Willis1999_S}
\begin{eqnarray}
\overline{q}_{11}(s) & = & -i\xi^-_c(s)+
\frac{i}{2} \xi^-_b(s)+   \left(\frac{a_+a_- -b_+b_-}{a_+a_- +b_+b_-}\right)+
\int_{C_-} \frac{d\xi_1 }{2\pi }\ln (T(s,\xi_1,1))  \label{q11}\ , \\
\overline{q}_{33}(s) & = & -i\xi^-_c(s)
+\frac{i}{2} \xi^-_a(s)  +\int_{C_-} \frac{d\xi_1}{2\pi } \ln (T(s,\xi_1,1))  \label{q33}\ ,
\end{eqnarray}
where
\begin{eqnarray}
T(s,\xi_1,1) &=&
\frac{\gamma^2 [ 4 \xi_{\bot}^2 \sqrt{(s-v\xi_1)^2/c_d^2-\xi_{\bot}^2} \sqrt{(s-v\xi_1)^2/c_s^2-\xi_{\bot}^2}
+((s-v\xi_1)^2/c_s^2-2\xi_{\bot}^2)^2 ]}{R(v)((s-v\xi_1)^2/c_R^2-\xi_{\bot}^2)(\xi_1-s/v)^2}\ ,\\
a_+ & = & \frac{i-\xi^-_c(s)}
{\sqrt{i-\xi^-_b(s)}} T_+(s,i,1) \;,\qquad a_- =  \frac{c_s^2 R(v) (-i-\xi^+_c(s) )}
{\beta v^2 \sqrt{-i-\xi^+_b(s)}} T_-(s,-i,1) \ ,\\
b_+ & = & \sqrt{i-\xi^-_b(s)} \;,\qquad b_-  =  \beta \sqrt{-i-\xi^+_b(s)}\ ,
\end{eqnarray}
with $\xi_{\bot}^2 \equiv \xi_1^2+1$ and
\begin{equation}
T_{\pm}(s,\pm i,1) =\exp \left( -\frac{1}{2\pi i} \int_{C_{\mp}} d\xi_1
\frac{\ln (T(s,\xi_1,1)}{\xi_1\mp i} \right)\ .
\end{equation}
The contours $C_{\pm}$ encircle the branch points $\xi_a^{\pm}$ and $\xi_b^{\pm}$ respectively, with
\begin{eqnarray}
\xi_a^{\pm} & = & -\frac{s \,v}{(c_d\alpha)^2} \pm i q_a; \qquad q_a  =  \frac{1}{\alpha}\sqrt{1-\frac{s^2}{(c_d \alpha)^2}} \ ,\\
\xi_b^{\pm} & = & -\frac{s \,v}{(c_s\beta)^2} \pm i q_b; \qquad q_b  =  \frac{1}{\beta}\sqrt{1-\frac{s^2}{(c_s \beta)^2}} \ ,\\
\xi_c^{\pm} & = & -\frac{s \,v}{(c_R\gamma)^2} \pm i q_c; \qquad q_c  =  \frac{1}{\gamma}\sqrt{1-\frac{s^2}{(c_R \gamma)^2}} \ ,
\end{eqnarray}
with $\alpha^2=1-v^2/c_d^2$, $\beta^2=1-v^2/c_s^2$, $\gamma^2=1-v^2/c_R^2$. The kinematic functions $\Theta_{13}(v)$ and $\omega_{13}(v)$ are defined in the main text.

Finally, using the homogeneity of the functions ${\bm U}$ and  $ \bm{P}^{(1)}$ and the time-independance of the zeroth order stresses, one can show that the function $T_{II}$ can be written in Fourier space as
\beq
\overline{T}_{II}(\xi_2,\omega)= K_I^{(0)}\sqrt{\frac{\left|\xi_2\right|}{L}}\,\overline{t}_{II}(\omega/\left|\xi_2\right|,\left|\xi_2\right|L)\,\overline{\psi^*}(\xi_2,\omega)\ ,
\eeq
where $\overline{t}_{II}$ is a {\it complex} function of degree zero and $L$ is the geometrical length scale introduced in the main text. Here we used the fact that to leading order in $\epsilon$ one has $\psi(X,x_2,t)= \psi(X+vt,x_2)$.

\section{Calculations}

For a real $s$ and $\sqrt{s^2+v^2}<c_R$, Eqs. (\ref{q11})-(\ref{q33}) reduce to
\bea
\Re\left[\overline{q}_{11}(s)\right] & = & - q_c
+\frac{1}{2}  q_b+   \left(\frac{a_+a_- -b_+b_-}{a_+a_- +b_+b_-}\right) +\Re\left[F_+(s)\right]\ ,\\
\Im\left[\overline{q}_{11}(s)\right] & = & \frac{s\,v}{(c_R\gamma)^2}
-\frac{1}{2} \frac{s\,v}{(c_s\beta)^2}   +\Im\left[F_+(s)\right] \label{im11} \ ,\\
\Re\left[\overline{q}_{33}(s)\right] & = & -q_c
+\frac{1}{2} q_a   +\Re\left[F_+(s)\right]  \ ,\\
\Im\left[\overline{q}_{33}(s)\right] & = & \frac{s\,v}{(c_R\gamma)^2}
-\frac{1}{2} \frac{s\,v}{(c_d\alpha)^2}  +\Im\left[F_+(s)\right]   \label{im33}\ ,
\eea
where
\beq
F_+(s)=\int_{C_-} \frac{d\xi_1 }{2\pi }\ln (T(s,\xi_1,1))\ ,
\eeq
The functions $a_+a_-$ and $b_+b_-$ are given by
\bea
b_+b_-&=&1+\sqrt{1-H^2/c_s^2}\ ,\\
a_+a_-&=&\frac{c_s^2R(v)\left(1+\sqrt{1-H^2/c_R^2}\right)^2}{v^2\gamma^2(1+\sqrt{1-H^2/c_s^2})}\exp[G_+(s)+G_-(s)]\ ,
\eea
with
\beq
H^2=s^2+v^2\ ,
\eeq
and
\beq
G_{\pm}(s)= - \int_{C_{\mp}}  \frac{d\xi}{2\pi i}
\frac{\ln (T(s,\xi,1)}{\xi\mp i}\ .
\eeq
The main calculations consist in computing the integrals $F_+(s)$ and $G_\pm (s)$. The algebraic manipulation of $F_+(s)$ was performed in~\cite{Ramanathan1997_S, Ramanathan1997a_S} and is given by
\bea
F_+(s)&=&\frac{1}{2\pi} \int_{c_s^2}^{c_d^2}  dJ \frac{2v^2J-H^2(J+v^2)}{\sqrt{J(J-H^2)}(J-v^2)^2} \tan^{-1}\left[\frac{4\sqrt{1-J/c_d^2}\sqrt{J/c_s^2-1}}{(2-J/c_s^2)^2}\right]\nonumber\\
&-& \frac{i}{\pi} \int_{c_s^2}^{c_d^2} dJ \frac{s\,v}{(J-v^2)^2} \tan^{-1}\left[\frac{4\sqrt{1-J/c_d^2}\sqrt{J/c_s^2-1}}{(2-J/c_s^2)^2}\right]
\label{Fplus}\ .
\eea
It is now established that the energy balance of the in-plane crack front dynamics leads to non-dispersive waves with real velocity $s_c$ for all crack front propagation speeds~\cite{Ramanathan1997a_S}. Using Eq.~(11) of the main text, one then has
\bea
\Re\left[\overline{q}_{33}(s_c)\right]&=&0
\label{re33sol}\ ,\\
\Im\left[2\overline{q}_{33}(s_c) - is_c \pa_v \left[\ln (\alpha (1-\beta^2)/R(v))\right]\right]&=&0
\label{im33sol}\ .
\eea
Eq.~(\ref{Fplus}) shows that $\Im[F_+]$ depends linearly on $s$. Consequently, the crack front wave solution $s_c$ is a solution of Eq.~(\ref{re33sol}) while Eq.~(\ref{im33sol}) is an identity which is satisfied for every real $s$ and $v$. This result can be checked through direct computation of the integral in~(\ref{im33sol}).
Moreover, one has
\beq
\Im\left[\overline{q}_{11}(s)\right] -\Im\left[\overline{q}_{33}(s)\right] =\frac{1}{2} \frac{s\,v}{(c_d\alpha)^2} -\frac{1}{2} \frac{s\,v}{(c_s\beta)^2} \ .
\eeq
Combining  this result with Eq.~(\ref{im33sol}) one gets
\beq
\Im\left[\overline{q}_{11}(s)\right] = \frac{s }{2}\pa_v \left[\ln (\beta (1-\beta^2)/R(v))\right] \ .
\eeq

In the following, we present the computation of $G_\pm (s)$ following the same steps as for the computation of $F_+$~\cite{Ramanathan1997_S}. Let us introduce the following change of variables in the above defined integrals
\beq
J=\left(\frac{s-v\xi}{\xi_{\bot}}\right)^2=\frac{(s-v\xi)^2}{1+\xi^2}\ .
\eeq
We then have
\beq
T(J)=\frac{\gamma^2v^2[(2-J/c_s^2)^2-4\sqrt{1-J/c_d^2}\sqrt{1-J/c_s^2}]}{R(v)J(J/c_R^2-1)}\ ,
\eeq
and
\beq
\xi_\pm(J)=\frac{-s\,v \pm i\sqrt{J(J-H^2)}}{J-v^2}\ ,
\eeq
leading to
\beq
J(\xi^{\pm}_a)=c_d^2\ , \qquad J(\xi^{\pm}_b)=c_s^2\ ,
\eeq
and
\beq
\xi'_\pm(J)=\frac{d\xi_{\pm}}{d J}=\frac{s\,v}{(J-v^2)^2}\pm \frac{i}{2}\frac{[H^2(J+v^2)-2v^2J]}{\sqrt{J(J-H^2)}(J-v^2)^2}\ .
\eeq
Note that we search for solutions for which $v^2<H^2<c_R^2$. Thus, the branch cut runs from $c_s$ to $c_d$ and the contribution of $\ln T(J)$  to the contour integral does not come from the denominator of $T(J)$ which is positive. Then, one has
\beq
G_{\pm}(s)= \frac{1}{\pi} \int_{c_s^2}^{c_d^2}  dJ \frac{\xi'_\mp(J)}{\xi_\mp(J)\mp i}\tan^{-1}\left[\frac{4\sqrt{1-J/c_d^2}\sqrt{J/c_s^2-1}}{(2-J/c_s^2)^2}\right]\ ,
\eeq
and thus
\beq
G_{+}(s)+G_{-}(s)= \frac{2}{\pi} \int_{c_s^2}^{c_d^2}  dJ \Re\left[\frac{\xi'_-(J)}{\xi_-(J)- i}\right]\tan^{-1}\left[\frac{4\sqrt{1-J/c_d^2}\sqrt{J/c_s^2-1}}{(2-J/c_s^2)^2}\right]\ ,
\eeq
with
\beq
2 \Re\left[\frac{\xi'_-(J)}{\xi_-(J)- i}\right]=-\frac{\left[2v^2(J+s^2)-H^2(J+v^2)\right]/(J-v^2)}{(s\,v)^2+(\sqrt{J(J-H^2)}+J-v^2)^2}
-\frac{\left[2v^2J-H^2(J+v^2)\right]/\sqrt{J(J-H^2)}}{(s\,v)^2+(\sqrt{J(J-H^2)}+J-v^2)^2}\ .
\eeq
With these algebraic manipulations and simplifications at hand, the numerical computation of $\Re [\overline{q}_{33}(s)]$ and  $\Re [\overline{q}_{11}(s)]$ is straightforward.

\end{document}